\documentclass[aps,prb,twocolumn,showpacs,floatfix]{revtex4-1}

\usepackage{longtable}

\usepackage[final]{graphicx}
\usepackage{dcolumn}
\usepackage{bm}

\usepackage{braket}
\usepackage{amsmath}

\usepackage{xcolor}

\newcommand{\mm}[1]     {\ifmmode {#1} \else{}${#1}$\fi}
\newcommand{\mmm}[1]    {\ifmmode{}#1 \else{}${#1}$\fi}
\newcommand{\beq}[1]{\begin{equation}\label{#1}}
\newcommand{\eeq}{\end{equation}}

\newcommand{\beqm}[1]{\begin{multiline}\label{#1}}
\newcommand{\eeqm}{\end{multiline}}

\def\vec#1{\mm{{\rm\bm{{\mathrm#1}}}}}

\def \mnge{{\rm MnGe}}
\def \mngem{{\rm MnGe$-$1}}
\def \mngef{{\rm MnGe$-$F}}

\def\P6m{\mm{P{6}/m}}

\begin{document}


\title{\large Topological magnetic structures in MnGe: Neutron diffraction and symmetry analysis}

\author{V.~Pomjakushin}
\affiliation{Laboratory for Neutron Scattering and Imaging, Paul Scherrer
Institut, CH-5232 Villigen PSI, Switzerland
}
\author{I.~Plokhikh}
\affiliation{Laboratory for Multiscale Materials Experiments, Paul Scherrer Institut, CH$-$5232 Villigen PSI, Switzerland
}
\author{J.S.~White}
\affiliation{Laboratory for Neutron Scattering and Imaging, Paul Scherrer
Institut, CH- 5232 Villigen PSI, Switzerland}
\author{Y. Fujishiro}
\affiliation{RIKEN Center for Emergent Matter Science (CEMS), Wako, Saitama 351-0198, Japan
}
\author{N. Kanazawa}
\affiliation{Department of Applied Physics, The University of Tokyo, Bunkyo-ku, Tokyo 113-8656, Japan
}
\author{Y. Tokura}
\affiliation{Department of Applied Physics, The University of Tokyo, Bunkyo-ku, Tokyo 113-8656, Japan
RIKEN Center for Emergent Matter Science (CEMS), Wako, Saitama 351-0198, Japan
}

\author{E.~Pomjakushina}
\affiliation{Laboratory for Multiscale Materials Experiments, Paul Scherrer Institut, CH-5232 Villigen PSI, Switzerland}

\date{\today}

\begin{abstract}

From new neutron powder diffraction experiments on the chiral cubic ($P2{_1}3$) magnet manganese germanide \mnge, we analyse all of the possible crystal symmetry-allowed magnetic superstructures that are determined successfully from the data. 
 The incommensurate propagation vectors $k$ of the magnetic structure are found to be aligned with the
 [100] cubic axes, and correspond to a magnetic periodicity of about 30 \AA\ at 1.8 K.
Several maximal crystallographic symmetry magnetic structures are found to fit the data equally well and are presented. These include topologically non-trivial magnetic hedgehog and ``skyrmion'' structures in multi-$k$ cubic or orthorhombic 3+3 and orthorhombic 3+2 dimensional magnetic superspace groups respectively, with either potentially responsible for topological Hall effect.
The presence of orthorhombic distortions in the space group $P2_12_12_1$ caused by the transition to the magnetically ordered state does not favour the cubic magnetic hedgehog structure, and leave both orthorhombic hedgehog and ``skyrmion'' models as equal candidates for the magnetic structures.
We also report on a new combined mechanochemical and solid-state chemical route to synthesise \mnge\ at ambient pressures and moderate temperatures, and compare with samples obtained by the traditional high pressure synthesis.

\end{abstract}


\maketitle


\section{Introduction}
Topologically nontrivial magnetic structures attract considerable interest because they can lead to new interesting phenomena like the topological Hall effect THE, which can be potentially useful for spintronic applications~\cite{Nagaosa}. Such structures are realised in the presence of more than one propagation vector of the magnetic structure. These propagation vectors should be symmetry related and reasonably small, so the magnetic textures based on the discrete localised magnetic moments are nanoscopically large. Typical examples are cubic MnSi \cite{doi:10.1126/science.1166767} or tetragonal CeAlGe \cite{puphal2020} with periodicity lengths of order 200~\AA\ and 70~\AA, and respectively hosting a lattice of magnetic particle-like objects called skyrmions or merons. In the latter case, the specific symmetry adapted magnetic structure in the magnetic superspace group (MSSG) was determined by neutron diffraction. In both cases the topological effects are revealed under external magnetic fields, but there is a fundamental difference in the field-evolution of the magnetic order. In CeAlGe there is no principal change in the magnetic structure, but in MnSi there seems to be an interesting change in the magnetic symmetry under the magnetic field as demonstrated from small angle neutron scattering SANS observations of the skyrmion structure \cite{doi:10.1126/science.1166767}. 

One more interesting example is cubic MnGe with magnetic periodicity about 30~\AA, which is suggested to display a topologically non-trivial ground state proposed in Refs.~\cite{kanazawa2021,kanazawa2012,Kanazawa_Nii2016,Kanazawa2017,Kanazawa_Kitaori2020} on the basis of SANS and electrical resistivity data. However, only a few wide-angle neutron diffraction studies are available, such as Ref.~\cite{kanazawa2021,mirebeau2012}. The magnetic structure was determined to be of a helical type with a propagation vector $\vec{k}$ = (0,0,$\delta$) in an orthorhombic symmetry $P2_12_12_1$~\cite{mirebeau2012}. 
This was the first neutron diffraction work, where the intensities of many magnetic and structural Bragg peaks were analysed, and it was suggested that the onset of the magnetic order coincides with a symmetry lowering. It was found that the magnetic modulation length falls with decreasing temperature and the structure possibly locks in to a commensurate one below 30 K ~\cite{mirebeau2012}.
From a further study of MnGe  by magnetic measurements, M\"{o}ssbauer spectroscopy, and neutron diffraction~\cite{mirebeau2014_3a} it was suggested that the zero field ground state at ambient pressure is a multidomain state consisting of helical domains with random orientations rather than a three-dimensional multi-\vec{k} lattice.
By combining resistivity, ac susceptibility, and neutron diffraction measurements under high pressure,  Ref.~\cite{mirebeau2014_3} showed that the helical order in MnGe transforms around 6~GPa from a high-spin to a low-spin state, recalling the weak ferromagnetism of MnSi at ambient pressure. 
The presence of several different magnetic-field-induced phases in MnGe was found from isothermal ac susceptibility experiments~\cite{Viennois_2015} presenting similarities with those in the isostructural compounds MnSi and FeGe but with much broader existence range of the A phase in the $(B,T)$ domain.
In Ref.~\cite{mirebeau2019}, using small-angle neutron scattering and a high-resolution method, the so-called MIEZE spectroscopy, it was shown that the proliferation of long-wavelength gapless spin fluctuations, concomitant with a continuous evolution of the helical correlation length appear upon cooling below the N\'{e}el temperature $T_N$ = 170 K. These fluctuations disappear at $T_{\rm com}$ = 32(5) K when the helical period becomes commensurate with the lattice.
This dynamic behaviour was in agreement with a previous local probe $\mu$SR study of MnGe~\cite{Andreica2016} where an inhomogeneous fluctuating chiral phase was found to set in with increasing temperature, characterized by two well-separated frequency ranges which coexist in the sample over a large temperature range below $T_N$.

In the context of determining the ground state magnetic structure of MnGe from scattering experiments, we think that a discrimination between a one-$\vec{k}$ modulated structure and multi-$\vec{k}$ topological magnetic structures is very difficult if not impossible from powder diffraction experiments due to the magnetic twin domains expected for the one-$\vec{k}$ structure. In some cases multi-$\vec{k}$ structures can be identified from the presence of specific higher order modulation harmonics in diffraction experiments~\cite{puphal2020}. Strong evidence in favour of different topological spin textures, such as skyrmion- and hedgehog-lattice states come from high-field transport measurements as demonstrated in the series of chiral magnets $\rm MnSi_{1-x}Ge_x$
~\cite{Fujishiro2019,kanazawa2022}.
Spin-polarized scanning tunneling microscopy, which is a direct space technique, was used to study surface magnetism in thin films of MnGe~\cite{Repicky}, revealing a variety of textures that are correlated to the atomic-scale structure. In contrast to bulk, the high spatial resolution images indicate three helical stripe domains and associated helimagnetic domain walls. 
Most notably, the hedgehog lattices parallel to the \{100\} atomic lattices were directly observed in MnGe also in real-space using high-resolution Lorentz transmission electron microscopy, simultaneously with underlying atomic-lattice fringes~\cite{Tanigaki2015}.
The inconsistency in the observed magnetic structures may stem from sample dependence. The formation of ultrashort-period magnetic structure in MnGe cannot be explained by the conventional model based on Dzyaloshinskii-Moriya interaction~\cite{c1,c2,Fujishiro2019}. In particular, recent theoretical studies~\cite{c4,c5,c6,c7} have revealed crucial roles of magnetic frustration and/or higher-order exchange interactions mediated by conduction electrons in the formation of short-period topological spin structures, including in the case of MnGe~\cite{c8,c9,c10}. Under this assumption, the magnetic structure would be highly dependent on the electronic state and would be therefore sensitive to crystallinity, compositional changes, strain and pressure; this is indeed supported by the observation of transitions among distinct topological spin crystals in $\rm MnSi_{1-x}Ge_x$~\cite{Fujishiro2019}. 

The motivation for the present work is to apply a state-of-the-art analysis of all possible magnetic superstructures allowed by the crystal symmetry in manganese germanide \mnge\ that are consistent with neutron diffraction data.
 The solution~\cite{mirebeau2012,mirebeau2014_3,mirebeau2014_3a} based on the single arm of the propagation vector $k$ seems to provide a good fit of the data, but is not actually unique. In addition, it cannot account for possible topological magnetic states. Here we present and analyse new solutions compatible with our powder neutron diffraction data, starting from maximal crystallographic symmetry magnetic structures for one $k$-vector (3D+1), three $k$-vectors (3D+3) and two $k$-vectors (3D+2) in 3D+n dimensional magnetic superspace groups (MSSG). The 3D+3 structure allows for topological hedgehog-type magnetic configurations consistent with those proposed in Ref.~\cite{kanazawa2012}.  

While single crystals of \mnge\ are not possible to grow at present in sizes suitable for neutron diffraction, we point out that if such single crystals were available, resolving between single-k and multi-k structures is still challenging due to the inevitable existence of magnetic twin domains. Even powder samples of \mnge\ have been hitherto difficult to synthesise due to necessity for high-pressure and temperature conditions. We have found a new route of sample synthesis at ambient pressure conditions and present it here as well.

\section{Sample synthesis. Experimental}
\label{exp}

Cubic phases of monogermanides CoGe, RhGe and MnGe with the $B$20-type structure are thermodynamically metastable under ambient conditions and can be synthesised at high pressures up to 8 GPa Refs.~\cite{Larchev1982,Takizawa1988}. One of our present samples, labeled as \mngef\ ($\sim$2g) was prepared by high pressure synthesis similarly as described in \cite{kanazawa2012} and the batch consisted of 9 samples, each made by individual high pressure syntheses. Here we have  also undertaken a new route to stabilise MnGe at ambient pressures. The sample labeled \mngem\ was prepared from the elements by a combined mechanochemical and solid-state route at ambient pressure. All preparations were performed in a He-filled grove-box (MBraun, $\rm O_{2}$ and $\rm H_{2}O$ less than 1~ppm). A stoichiometric mixture (1.5g) of metallic Mn (over 99.9\%, purified from surface oxide in $\rm HNO_{3}$) and Ge (over 99.9\% purity) was treated mechanochemically (100 rpm, 5 min pre-milling, +2 min cooling, 2 cycles, 600rpm, 5 min milling, +5 min cooling, 10 cycles) in a Pulverisette p-7 ball-mill (Fritsch, Germany). The obtained powder was pressed into a pellet, flame-sealed in evacuated quartz ampoule and annealed at 400$^{\circ}$C for 2 days. The laboratory powder X-ray diffraction of the sample has shown a presence of the main phase (MnGe) with a pronounced amount of impurity phases (starting elements (Mn, Ge) and some intermediate phases). The sample was thus retreated mechanochemically again (800 rpm, 5 min milling, +5 min cooling, 10 cycles) and no impurity phases were detected with powder XRD.

Neutron powder diffraction experiments on both samples were carried out at the SINQ spallation source at the Paul Scherrer Institute (Switzerland) using the high-resolution diffractometer for thermal neutrons HRPT \cite{hrpt} 
using wavelengths $\lambda=1.494$~\AA\ and 2.45\AA\ and different modes of operation: high intensity HI for magnetic diffraction, medium resolution MR and high resolution HR with resolutions $\delta d/d >1.8\cdot 10^{-3}$, $>1.3\cdot 10^{-3}$ and $>0.9\cdot 10^{-3}$, respectively,  achieved by the primary white beam collimations 40', 12' and 6'\cite{hrpt}. The intensities in the MR and HR modes are significantly smaller, amounting 30\% and 7\% of the HI-mode, but their use allowed us to resolve the fine details of the crystal structure. 
The determination of the crystal and magnetic structure parameters were done using the {\tt FULLPROF}~\cite{Fullprof} program, with the use of its internal tables for neutron scattering lengths. The symmetry analysis was performed using {\tt ISODISTORT} from the {\tt ISOTROPY} software \cite{isod,isod2} and some software tools of the Bilbao crystallographic server such as  {\tt MVISUALIZE} \cite{Bilbao,ISI:000358484200010}.

\section{Crystal structure and microstructure}
\label{str}
The crystal structures in both samples are well refined in the cubic space group $P2_1 3$ (no. 198) with the following structure parameters at 1.8~K. Both atoms are  in  $4 a$ positions $(x,x,x)$,  with  $x_{\rm Ge}$ = $0.15678(17)$,  $x_{\rm Mn}$ = $0.8620(4)$,  lattice constant $a= 4.7782(4)$\AA\ for \mngem, and $x_{\rm Ge}$ = $0.1568(2)$, $x_{\rm Mn}$ =  $0.8639(5)$, $a=4.7805(2)$\AA\ for \mngef. The illustrations of the refinement quality are shown in Fig.~\ref{rt_diff}. The coherently scattering domains (or crystalline sizes) are relatively small for the \mngem\ sample, amounting to $L=150$~\AA.  One can see this effect by inspection as
 diffraction peak broadening. In the \mngef\ sample the  peaks are narrow, implying large ($>2000$\AA) crystalline sizes. The small crystallines in the \mngem-sample are thus apparently due to the synthesis technique.
We point out that the upturn of the intensity profile (Fig.~\ref{rt_diff}) towards $2\theta\rightarrow 0$ is due to the short range magnetic correlations, and not due to instrumental background.

On cooling below $T_N$ we observe additional Bragg peak broadening in the \mngef\ sample due the microstrain effect $\delta a/a$, caused by either the distribution of the lattice constant size $a$ over different crystallines or crystal domain or due to a lowering of the symmetry below cubic. Due to the high resolution and large Q-range at HRPT, we can distinguish between the effect of microstrains and the finite size effects. 
We also performed the measurements using HR and MR modes of HRPT to better determine the origin of the peak broadening effect.  We further made two comparative fits of MR- and HR-datasets: firstly at 2~K in the cubic model with microstrains and secondly in the orthorhombic space subgroup $P2_12_12_1$ (no. 19) with the microstrains fixed to those refined from the 300~K pattern with all structure parameters and crystal metric released. The average microstrain in cubic model amounted to $4.7\cdot 10^{-4}$ and $1.8\cdot 10^{-4}$ at T=2~K and 300~K, respectively. The HR-mode allows us to unambiguously determine a preference for the orthorhombic symmetry. In addition to an overall improvement of the well convergent fit in the $P2_12_12_1$ model, one can see the asymmetric shoulders and the peak width misfit in the cubic model for Bragg peaks located two-theta region of highest resolution, as shown in Figure \ref{HR}. The diffraction peak (320) has a clearly visible shoulder towards high angle, which is ideally fit by the orthorhombic peak splitting. The  Bragg peak $(222)$ is not split in both groups, but it is badly fit in the cubic model with microstrains due to wrong peak width.
The fit of the combined  MR- and HR-datasets using both wavelengths in the orthorhombic subgroup converged with the following structure parameters  Ge ($4a$): [0.1549(6), 0.1581(8), 0.1569(8)], Mn ($4a$): [0.861(1)  0.865(1), 0.862(1)] and the cell parameters $a, b, c$ =  4.7776(1), 4.7823(1), 4.7854(1)~\AA.
We note that the use of the HR-mode was crucial for finding the orthorhombic distortions. This is due to the fortunate circumstance that the instrumental resolution at the position of the (320)-peak at $2\theta=135$~degrees (for $\lambda=2.45$~\AA) is 50\% higher than in MR-mode, allowing us to distinguish peak broadening from peak splitting. Naturally the splitting is present in many Bragg peaks, but not so explicitly asymmetric as in the above mentioned peak. 
Since the magnetic structure models (as is discussed in section \ref{symmetry}) for three k-vectors allow also the rhombohedral crystal structure subgroup $R3$ (no. 146) we have also considered this structure model, which gives rise to the splitting of the cubic peaks as well. Contrary to the orthorhombic model the $R3$ crystal metrics fails to describe the experimental peak shapes and thus the rhombohedral solution can be excluded (some details and the illustration of the fit quality are shown in Fig.~SM2).

\section{Magnetic structure determination and discussions}
\label{mag_str}
The magnetic diffraction patterns are hallmarked by a very large first diffraction peak, this being the so called zero satellite $(0,0,0)\pm(0,b,0)$. Remarkably, they are even more intense than the nuclear Bragg peaks. In contrast, the majority  of the other magnetic peaks at larger scattering angles
 have relatively small intensities, as one can see in Figures \ref{mag_diff_01} and \ref{mag_diff_F}.  For this reason, difference patterns, i.e. the difference between patterns taken at base and paramagnetic temperatures, were used to solve and refine the magnetic structure. Such difference patterns contain purely magnetic scattering and are free of possible systematic uncertainties due to the fitting
 of large crystal structure Bragg peaks, background, impurities, etc. There are two difficulties that make the subtraction of patterns not completely straightforward, namely the high magnetic transition temperature and thermal variation of the crystal structure results
in a substantial difference in the lattice constant $a$, and the presence of strong short range scattering just above the Ne\`{e}l temperature $T_N\simeq170$~K. We used the paramagnetic pattern measured at highest temperature 300~K, where the short range ordering effects are smallest. To compensate for the difference in the lattice constants at the different temperatures, we
 have fitted the 300~K pattern with the crystal metric fixed by the value determined at base temperature, and instead refining
 the neutron wavelength. Then we recalibrated the diffraction pattern at 300K with the refined wavelength before performing the subtraction, thus resulting in
 a very good difference pattern. 
 

We have further tried to estimate the magnetic short range correlation length at T=300~K from the low angle scattering (Fig.~\ref{rt_diff}) for \mngef\ sample, which is mainly from the tail of the diffuse magnetic Bragg peak. We subtracted the direct beam contribution which amounts to maximum 20\% of the diffuse intensity at two theta $2\theta>4^\circ$ and performed the fit to pseudo-Voigt function in the range between 4 and 20 degrees. The full width half maximum (FWHM) and the peak position were refined to be 6.2 and 2.7 degrees, respectively. The peak position corresponds to the propagation vector length $k=0.093$ r.l.u, which is in good agreement with its value of 0.11 at 160~K (the length $k$ substantially falls as the temperature approaches $T_N$)~\cite{mirebeau2012}. The peak broadening and the position correspond to a correlation length and modulation period about $L\simeq20$~\AA\ and $t_{\rm mag}\simeq50$~\AA\ respectively. It interesting to compare these values with the ones for the long range ordering at base temperature  2~K, which are about $L\simeq 520$\AA\ and 30\AA, respectively. 


\subsection{Model free Le Bail fit}
The identification of the magnetic propagation vectors was done using so called Le~Bail fitting,  where all peak intensities are refined separately without any structure model, thus allowing a straightforward determination of the 
 the propagation vectors $k$ and the crystal metrics. Both samples display the same type of propagation $[0,b,0]$, which is the delta point DT  of the Brillouin zone BZ [here we use internationally established nomenclature for the irreducible representations (irreps) labels  and magnetic superspace groups MSSG~\cite{Bilbao,isod}].  The refined values amounted to $b=0.17395(5)$ and $0.16498(3)$  in reciprocal lattice units for
the \mngem\ and \mngef\ samples respectively. All peaks in the difference magnetic patterns could be indexed with the single propagation vector $[0,b,0]$. One can use any of three nonequivalent k-vectors for the indexing; here we use the $y$-direction in accord with the established nomenclature.
The total number of             
 independent reflections is 36, with 16 among them scattering at non-degenerate scattering angles in two theta.

\subsection{Symmetry analysis and magnetic models}
\label{symmetry}

The parent space group $P2_13$ (no. 198) has two irreps for the delta point DT $(0,b,0)$ of the BZ.  The irrep mDT1 does not describe the data at all, because it predicts zero intensity for the most intense first magnetic Bragg peak. So the solution is irrep mDT2, which results in three maximal symmetry MSSG. According to the cubic symmetry we have three  models based on a single arm (3+1), two arms (3+2) and three arms (3+3) of the propagation vector star. In each model the Mn-atom remains unsplit and retains the
 single ($4a$) position. 

First we consider two most symmetric magnetic models based on one (3+1) and three (3+3) arms of the propagation vector. The single $k$-vector model corresponds to the MSSG 19.1.9.1.m26.2 P2\_12\_12\_1.1'(0b0)0s0s, whereas for the 3+3 model, MSSG is 198.3.206.1.m10.2 P2\_13.1'(a,0,0)00s(0,a,0)00s(0,0,a)00s. 
Both MSSGs allow 6 free parameters describing the amplitudes of
 cosine and sine components for $x$-, $y$- and $z$-components of the magnetic moment. To avoid ambiguity in the description of the magnetic configuration in a MSSG\footnote {Depending on the basis transformation from the parent group and propagation vector choice one can have different symmetry operators. They are listed in supplementary materials together with magnetic crystallographic information files (mcif).}, 
below we list explicitly the 3D-symmetry operators, the Mn-coordinates and the formulae for the magnetic moments.

It the 3+1 model the internal coordinate is $x_4=(\vec{k}_1\cdot \vec{X})$, where  $\vec{X}$ is the fractional coordinate of respective Mn-atom in position $\vec{X}$,  and $\vec{k}_1=(0,b,0)$ is the propagation vector. In crystallographic notation, one usually uses  sine and cosine components of the magnetic moment propagation and $x_4, x_5, x_6$ internal coordinates. Here, for brevity we use a cosine modulation with the amplitudes $m$ and phases $\alpha$,  and the reduced spacial coordinate  $\tilde{y}=2\pi x_4$.  Formula (\ref{mod1}) shows the explicit form of the modulation for four Mn-moments. The moment components are surrounded by square brackets for each Mn-site from one to four. Note, that the relations between the signs of $m$ and $\alpha$ for the different Mn-positions are dictated by the magnetic symmetry.  The notation for the positions is the following: Mn1 x,y,z (0.8616, 0.8616, 0.8616), Mn2 -x+1/2,-y,z+1/2 (0.6384, 0.1384, 0.3616), Mn3 -x,y+1/2,-z+1/2 (0.13840, 0.3616, 0.6384), Mn4 x+1/2,-y+1/2,-z (0.3616, 0.6384, 0.1384). 
%
\newcommand\numberthis{\addtocounter{equation}{1}\tag{\theequation}}
 \begin{align*}
 M_1[ m_{1} \cos(\tilde{y} +\alpha_1)  ,
 m_{2} \cos(\tilde{y} +\alpha_2)  ,
 m_{3} \cos(\tilde{y} +\alpha_3)  ]\notag\\
 M_2 [ m_{1} \cos(\tilde{y} -\alpha_1)  ,
 m_{2} \cos(\tilde{y} -\alpha_2)  ,
 -m_{3} \cos(\tilde{y} -\alpha_3)  ]\notag\\
  M_3 [ m_{1} \cos(\tilde{y} +\alpha_1)  ,
 -m_{2} \cos(\tilde{y} +\alpha_2)  ,
 m_{3} \cos(\tilde{y} +\alpha_3)  ]\notag\\
 M_4[ m_{1} \cos(\tilde{y} -\alpha_1)  ,
 -m_{2} \cos(\tilde{y} -\alpha_2)  ,
 -m_{3} \cos(\tilde{y} -\alpha_3)  ]\notag\numberthis{\label{mod1}}
 \end{align*}
The 3+3 model has 6 independent parameters, as well. The internal coordinates are $x_4=(\vec{k}_1\cdot \vec{X})$, $x_5=(\vec{k}_2\cdot \vec{X})$ and  $x_6=(\vec{k}_3\cdot \vec{X})$ where  $\vec{X}$ is the fractional coordinate of respective Mn-atom, $\vec{k}_1=(0,b,0)$,  $\vec{k}_2=(0,0,b)$ and $\vec{k}_3=(b,0,0)$. The reduced spacial coordinates  are $\tilde{y}=2\pi x_4$, $\tilde{z}=2\pi x_5$ and $\tilde{x}=2\pi x_6$. Formula (\ref{mod3}) shows the explicit form of the modulation for the four Mn-moments. 
In spite of the fact that there are three independent $k$-vectors, the amplitudes and phases on different arms are constrained by the high cubic symmetry. 
 \begin{align*}
 M_1[ m_{1} \cos(\tilde{y} +\alpha_1) + m_{3} \cos(\tilde{z} +\alpha_3) + m_{2} \cos(\tilde{x} +\alpha_2)  ,\\
 m_{2} \cos(\tilde{y} +\alpha_2) + m_{1} \cos(\tilde{z} +\alpha_1) + m_{3} \cos(\tilde{x} +\alpha_3)  ,\\
 m_{3} \cos(\tilde{y} +\alpha_3) + m_{2} \cos(\tilde{z} +\alpha_2) + m_{1} \cos(\tilde{x} +\alpha_1)  ]\notag\\
 M_2[ m_{1} \cos(\tilde{y} -\alpha_1) + m_{3} \cos(\tilde{z} +\alpha_3) -m_{2} \cos(\tilde{x} -\alpha_2)  ,\\
 m_{2} \cos(\tilde{y} -\alpha_2) + m_{1} \cos(\tilde{z} +\alpha_1) -m_{3} \cos(\tilde{x} -\alpha_3)  ,\\
 -m_{3} \cos(\tilde{y} -\alpha_3) -m_{2} \cos(\tilde{z} +\alpha_2) + m_{1} \cos(\tilde{x} -\alpha_1)  ]\\
 M_3[ m_{1} \cos(\tilde{y} +\alpha_1) -m_{3} \cos(\tilde{z} -\alpha_3) + m_{2} \cos(\tilde{x} -\alpha_2)  ,\\
 -m_{2} \cos(\tilde{y} +\alpha_2) + m_{1} \cos(\tilde{z} -\alpha_1) -m_{3} \cos(\tilde{x} -\alpha_3)  ,\\
 m_{3} \cos(\tilde{y} +\alpha_3) -m_{2} \cos(\tilde{z} -\alpha_2) + m_{1} \cos(\tilde{x} -\alpha_1)  ]\notag\\
 M_4[ m_{1} \cos(\tilde{y} -\alpha_1) -m_{3} \cos(\tilde{z} -\alpha_3) -m_{2} \cos(\tilde{x} +\alpha_2)  ,\\
 -m_{2} \cos(\tilde{y} -\alpha_2) + m_{1} \cos(\tilde{z} -\alpha_1) + m_{3} \cos(\tilde{x} +\alpha_3)  ,\\
 -m_{3} \cos(\tilde{y} -\alpha_3) + m_{2} \cos(\tilde{z} -\alpha_2) + m_{1} \cos(\tilde{x} +\alpha_1)  ]\notag\\
\numberthis{\label{mod3}}
 \end{align*}
%
%
The 3+2 model is based on arms of the propagation vector star
 with MSSG 19.2.29.2.m26.3 P2\_12\_12\_1.1'(0,b1,0)000s(0,0,g2)000s and has the same orthorhombic symmetry as the 3+1 model, instead allows for 12 independent parameters, because the moments propagating by different arms are not symmetry related. The internal coordinates are $x_4=(\vec{k}_1\cdot \vec{X})$, $x_5=(\vec{k}_2\cdot \vec{X})$, where  $\vec{X}$ is the fractional coordinate of the respective Mn-atom, $\vec{k}_1=(0,b,0)$ and $\vec{k}_2=(b,0,0)$. 
Formula (\ref{mod2}) shows the explicit form of modulation for the four Mn-moments.  Note that the magnetic configuration based only on the first arm $\vec{k}_1$ is identical to one given by the 3+1 model (\ref{mod1}). This model is interesting, because as we show below, it allows ``skyrmion'' type of magnetic structure. We label the structure as the ``skyrmion''-type, following the Ref.~\cite{kanazawa2012}, because it is propagating in 2D-plane to distinguish it from the cubic hedgehog structure. As we show below, the ``skyrmion''-structure hosts the partical-like objects that can be identified as merons and antimerons.
 \begin{align*} 
M_1=[ m_{1} \cos(\tilde{y} +\alpha_1) + m_{4} \cos(\tilde{x} +\alpha_4)  ,\\
 m_{2} \cos(\tilde{y} +\alpha_2) + m_{5} \cos(\tilde{x} +\alpha_5)  ,\\
 m_{3} \cos(\tilde{y} +\alpha_3) + m_{6} \cos(\tilde{x} +\alpha_6)  ]\notag\\
M_2=[ m_{1} \cos(\tilde{y} -\alpha_1) -m_{4} \cos(\tilde{x} -\alpha_4)  ,\\
 m_{2} \cos(\tilde{y} -\alpha_2) -m_{5} \cos(\tilde{x} -\alpha_5)  ,\\
 -m_{3} \cos(\tilde{y} -\alpha_3) + m_{6} \cos(\tilde{x} -\alpha_6)  ]\notag\\
M_3=[ m_{1} \cos(\tilde{y} +\alpha_1) + m_{4} \cos(\tilde{x} -\alpha_4)  ,\\
 -m_{2} \cos(\tilde{y} +\alpha_2) -m_{5} \cos(\tilde{x} -\alpha_5)  ,\\
 m_{3} \cos(\tilde{y} +\alpha_3) + m_{6} \cos(\tilde{x} -\alpha_6)  ]\notag\\
M_4=[ m_{1} \cos(\tilde{y} -\alpha_1) -m_{4} \cos(\tilde{x} +\alpha_4)  ,\\
 -m_{2} \cos(\tilde{y} -\alpha_2) + m_{5} \cos(\tilde{x} +\alpha_5)  ,\\
 -m_{3} \cos(\tilde{y} -\alpha_3) + m_{6} \cos(\tilde{x} +\alpha_6)  ]\notag\\
 \numberthis{\label{mod2}}
 \end{align*}
%
%
Finally we present two 3+3 models with the symmetry lower than cubic, both having 18 independent parameters, because the moments propagating by different arms are not symmetry related, similarly as for the above 3+2 model. The model 3+3O with MSSG 19.3.95.4.m26.4 P2\_12\_12\_1.1'(a1,0,0)000s(0,b2,0)000s(0,0,g3)000s has the same orthorhombic symmetry as the 3+1 and 3+2 models. In the 3+3O orthorhombic model the modulations along three directions $\tilde{y}$, $\tilde{z}$ and $\tilde{x}$ in formula (\ref{mod3}) are no longer symmetry related, but the 3+3 model given by (\ref{mod3}) is a valid partial solution of this less symmetric subgroup of the 3+3 cubic MSSG. The second 3+3 subgroup is MSSG 146.3.185.3.m11.2 R3.1'(a,b,g)0s(-a-b,a,g)0s(b,-a-b,g)0s. This rhombohedral magnetic model can be excluded from the consideration because the high resolution diffraction patterns are not compatible with the rhombohedral distortions as discussed previously in section \ref{str}.

\subsection{A note on a continuous limit of the magnetic structure for the different
 MSSG}
\label{anote}
The magnetic structure is defined on the discrete set of points $\vec{r}_j$ given by the positions of atoms in the
 crystal lattice. In the case of an incommensurate structure, 
 the size and direction of the atomic magnetic moments related by the propagation $\vec{k}$-vector are proportional to  $\cos(\vec{k}\cdot\vec{r}_j)$, and in the limit of $\vec{k}\to0$  one can approximate the distribution of the magnetization density to be spatially continuous.
 However, there is a principal difficulty in the realisation of the continuous limit related to the crystallographic symmetries.  
In general, due to the specific space group symmetry one has several atoms in the primitive unit cell (except when the multiplicity is 1). The spins of these atoms are related to each other by crystallographic operators like rotations by the large crystallographic angles, such as 180, 120, 90  or 60 degrees. So even in the limit of $\vec{k}\to0$ we might have a finite (not going to zero) angle between the spins of neighbouring atoms. These rotations can be compensated to some extent by the irrep-matrices in some cases. In the present case we have four atoms related by 180 degrees rotations along x, y and z-axis, but the mDT2-irrep matrices are constructed
 so that the moments are rotated back 
for some specific moments directions. For instance, for the 3+3 model, and for only nonzero $m_1$ and small $\alpha_1$, the rotations are practically compensated, as seen from formula (\ref{mod3}). However, if $\alpha_1=\pi/2$ then the x-component changes direction to be opposite between neighboring Mn-atoms thus making continuous limit of $\vec{k}\to0$ impossible. 

Interestingly, the constraints that we have by symmetry in 3+3 cubic MSSG, and the partly symmetry related constraints we can apply in 3+2 orthorhombic MSSG, allow the hedgehog-type and skyrmion-type structures, respectively. Both structures are compatible with the
 continuous limit. 
 
 For the 3+3 model, if we put $m_1=m_3=1, \alpha_3=\pi/2 $ and the other parameters to zero, then we get the following components of magnetisation:
\beq{hedgehog}
(m_x, m_y, m_z) = (\cos{y} - \sin z, \cos z - \sin x, -\sin y + \cos x). 
\eeq
With this parametrisation the 180 degrees rotations are completely compensated by the phases, as one can see from formula (\ref{mod3}), leaving only changes in the moment given by propagation vectors. We denote this structure as 3+3 hedgehog and it is similar to one proposed in Ref.~\cite{kanazawa2012}. 

For the orthorhombic 3+2 model, one can choose $m_1=m_3=m_5=m_6=1$ and $\alpha_3=\alpha_5=\pi/2$ in (\ref{mod2}) and also have the compensation of the rotations between neighbouring moments, leading to the following skyrmion like magnetisation:
\beq{skyrmion}
(m_x, m_y, m_z) = (\cos{y}, -\sin x, -\sin y + \cos x). 
\eeq

%

In the strict sense, the absence of the continuous limit in the general case of high symmetry space groups is contradictory to the notion of topological non-triviality.
One can think that the conduction electrons following the local magnetization adiabatically contribute to
 a continuous distribution of the magnetisation $M(\vec{r})$. However, then the functional form of $M(\vec{r})$ between the rotated moments is non-harmonical.  To get the possibility that in the general case all four Mn moments follow the same cosine and sine modulations, one should remove all symmetry restrictions imposed by the crystal symmetry. For the 3+3 magnetic structure this will lead us to the lowest symmetry triclinic MSSG group  1.3.1.1.m2.2 P1.1'(a1,b1,g1)0s(a2,b2,g2)00(a3,b3,g3)00 with 18 free parameters: three cosine and three sine components [or three amplitudes $m$ and three phases $\alpha$ in formula (\ref{mod3})] for each $k$-vector. Technically, in formula (\ref{mod3}) it is necessary
 to keep the same amplitudes and phases separately for each component propagating as $\tilde{x}$, $\tilde{y}$ and $\tilde{z}$ without constraints between them. The expressions for Mn2, Mn3 and Mn4 should be identical to the one for Mn1.

\subsection{Experimentally confirmed magnetic structures}
\label{magstr}
First, using the {\tt FULLPROF} program, we
 have performed a simulated annealing (SA) minimization \cite{kirkpatrick83,Fullprof} of the full diffraction profile, containing 36 magnetic Bragg peaks for the models described in the previous sections.  A SA search  starts from random values of the free parameters and we have repeated the search more than several hundreds times. The reliability profile factors $Rp$ (in percents) for the solutions came in the ranges 2.44~-~2.5, 2.46 - 2.56 and 2.45 - 2.46 for the 3+1, 3+3 and 3+2 magnetic models, respectively. 
%
%
The searches converged to one or two solutions for the 3+1 and 3+3 models respectively, in
 the ranges of $Rp$ shown above. Finally the result of the SA search was refined further using a usual least square Rietveld refinement of the powder diffraction pattern. The goodnesses of final fits were similar to those from Le Bail fitting which had chi-square $\chi^2\simeq 4$, implying that there is no room for further improvement.
%

%
%
The results of the fits are summarised in tables \ref{magmom1},\ref{magmom3}. The very strong zero satellite with a difficult to handle asymmetric peak shape provides the main contribution to the chi-square $\chi^2$. This peak is important for the data analysis, but if we exclude it from the fit then the $\chi^2$ falls from 5 to about 1, implying that all fits given in the tables are very good.

For the 3+1 model the SA search finds two types of models: an amplitude modulated (AM) structure and approximately helically modulated structure (helix) with insignificantly slightly larger $Rp$. They are both listed in the table. The helix-model has large moment components perpendicular to the propagation vector $[0,b,0]$ with cosine and sine modulations, whereas the AM-structure does not have a sine-contribution for the respective perpendicular component. The models after SA have redundant number of parameters but we present them for the completeness to show the largest components of the moments.
The quality of fits is illustrated in Figures \ref{mag_diff_01}, \ref{mag_diff_F}, where the experimental and calculated diffraction patterns are shown.  In Table \ref{magmom1}, we show the
 three minimal 3+1 models that equally well fit the data: (i) with cosine and sine component along the same $x$-axis (AM-x), (ii) with cosine modulations along both $x$ and $z$ axes (AM-xz), and (iii) the helical modulation structure. 
The helix-structure has only one parameter and has a similar goodness of fit as
 for the AM structures. Note, that the modulation amplitude for the helical model is naturally about $\sqrt{2}$-times smaller due to the constant moment structure. For the comparison between models it is better to use $\chi^2$ than the conventional $R$-Bragg because the latter does not take into account experimental errorbars, and due to the presence of the very strong zero satellite might be slightly misleading. In any case, the $R$-factors are very good, being 0.5\% for \mngef\ sample. The illustrations of the magnetic structures are shown in Figure~SM2.

%

For the 3+3 model the SA-search converges to the single hedgehog-type structure described in section \ref{anote}. The results of the SA-search and subsequent conventional Rietveld fits are shown in Table~\ref{magmom3}. In the final least square fit we restricted the components along x- and z-axis to be the same leaving only a single fit parameter $m_{xs}=-m_{zc}$. The hedgehog structure is truly continuous in the limit of $\vec{k}\to0$  with the magnetisation distribution given by formula (\ref{hedgehog}).
In addition to the most symmetric hedgehog model, we present two solutions, which fit the data equally well: (i) with the components along x and (ii) with cosine component along x and components along z, denoted in the same way as for 3+1 case. This 3+3 model has highest possible cubic symmetry and fits the data with the similar goodness of fit as 3+1 model. As discussed in section \ref{anote}, these solutions are quasi-continuous, because the crystallographic rotations are practically compensated for the sets of parameters found to fit the data well.

According to the refined magnetic structures, the magnetic
 moments propagate along all three cubic axes, thus creating rather complicated distribution that is difficult to illustrate in all three dimensions. In Figures~\ref{mag_str_F3d4x4x1}, \ref{mag_str_F3d11x11x1} we attempt to present several views of the hedgehog structure on the microscopic lattice.
Figure~SM3 
shows the first 15x15x15 unit cells in projection along (111) and (100) cubic axes nicely demonstrating six- and four-fold textures. 

The 3+2 model can definitely fit the data because the equations (\ref{mod2}) for the first arm of the propagation vector star are identical to the 3+1 model. The  SA-search for the 3+2 model with 12 independent parameters finds several solutions that have the same goodness in $Rp$. Among them there is the minimal most symmetric ``skyrmion'' model corresponding to the formula (\ref{skyrmion}). The results of the fit are given in the Table \ref{magmom3}. The second arm parameters are constrained to the first arm as explained in section \ref{anote}. The magnetic structure is illustrated in Fig.~\ref{mag_str_F2d11x11x1} and hosts the partical-like objects that can be identified as merons and antimerons (see section \ref{topoQ}).


Both the hedgehog 3+3 cubic model and the 3+2 model fit the data equally well. However, as shown above in section \ref{str}, the crystal structure is orthorhombic with the space group $P2_12_12_1$. The hedgehog structure is orthorhombically distorted and corresponds to the 3+3O model with partial symmetric solution given by 3+3 cubic group. The beauty of the cubic solution is that the hedgehog configuration of the magnetic moments is dictated by the maximal symmetry. For the 3+3O model the moment amplitudes for all four atoms propagating along each arm are still fixed by the symmetry in the same way as in the 3+3 model, but the symmetry relations between the arms are lost due to the lack of 3-fold axis.  The orthorhombic distortions are very small, and any deviations of the magnetic structure from the cubic model are not possible to resolve experimentally.  Due to the difference in the lattice constants, the modulation periods of the hedgehog structure will be different as well, with relative difference in periods of 0.1\% and 0.16\% along the $b$ and $c$ axes with respect to the $a$-axis following the crystal metrics determined in section \ref{str}. In view of the fact that the deviations from cubic metrics are so small, the cubic model might be still a good approximation for the exchange interactions in this system.
Independent evidence for the 3+3 hedgehog structure come from the high-resolution Lorentz transmission electron microscopy \cite{Tanigaki2015} and  the SANS studies on single-crystalline MnGe thick films~\cite{Kanazawa2017,Kanazawa2020}.

Both 3$k$-cubic hedgehog and 2$k$-othorhombic ``skyrmion'' structures can be responsible for the topological hall effect observed in \mnge~\cite{kanazawa2021,kanazawa2012} and should be thus preferred
 over the simple one-$k$ helical structure. We like to point out that the topological structures are partial solutions in the indicated MSSG, whereas the general solutions do not have continuous limit structures. The hedgehog and ``skyrmion'' structures can be viewed as a sum of 3 or 2 helical single-$k$ structures, resulting in modulated structures with non constant magnetic moment.

\section{Topological charges of magnetic structures}
\label{topoQ}

In the continuous limit of the 3+2 model for $k\to0$ one can readily calculate the density of the topological index as
\beq{top_idx}
w(x,y)={1\over4\pi} (\vec{n} \cdot [\frac{\partial \vec{n}}{\partial x}\times \frac{\partial \vec{n}}{\partial y} ]),
\eeq
where $\vec{n}=\vec{m}/|{\vec{m}|}$, and $\vec{m}(x,y)$ are the functions for the magnetic moment components given by formula (\ref{skyrmion}).  In zero field there are two singularities per magnetic cell located at coordinates (0,$3\pi\over2$) and ($\pi$,$\pi\over2$), where all three components of the magnetisation $\vec{n}$ become zero. To avoid singularities in calculations and visualisations the coefficient of the cosine for z-component was chosen to be 1.0001 instead of 1. 
The maxima and minima
 of $w(x,y)$ look like localised particle-like objects with topological charge $Q=\pm1/2$, where $Q= \int w(x,y) dx dy$. 
In infinitesimally small magnetic field along the z-axis (ferromagnetic $\pm m_f$ component added to z-component of magnetisation) each peak acquires the same charge $Q=\mp 1/2$, as shown in Fig.~\ref{topo_den_mf} making in total skyrmion-like charge $Q=\mp1$ per unit cell. But the fundamental magnetic objects themselves are not skyrmions, but merons.
 The total charge maintains an integer value $Q=-1$ until $m_f$ reaches the critical value $m_c=2$ , above which the total charge becomes abruptly zero. The magnetic structure is not yet fully ferromagnetic (FM) polarised, but since the antiferromagnetic amplitude is $M= 2$ the moment values are always larger than zero. Figure~\ref{topo_den_mf} illustrates the evolution of $w(x,y)$ as a function of field. Interestingly, at intermediate field $m_f=1.0$ the sharp minima change to smoother distribution of density and then closer to the limit there is only one peak carrying most of the charge $Q=-1$. In the FM polarised state there are some sharp features like at $m_f=2.1$, but the total charge is $Q=0$. For the field directed away from the z-axis, the upper field $m_f$, when the charge becomes zero, is smaller than $m_c=2$.  A simple powder averaging of the critical field $m_c$, above which the charge $Q$ is zero gives $\left<m_c\right>$=1.6.

This is an oversimplified toy model, because it does not assume any fundamental change in the magnetic structure over the full range of the magnetic field up to FM saturation state - we only add a FM component to the rigid magnetic configuration. Such a simple approach appeared to work
 in the case of small external magnetic fields in the range where the THE was observed in CeAlGe \cite{puphal2020}. 
In a more realistic model one can consider canting of the moments toward the field, but since only the normalised magnetisation is used for the calculation of charge density (so that only the direction is important) we think that the toy model may capture  qualitative aspects of the field-dependent behaviour.
At high fields the saturated magnetic moment should be approximately the
 square averaged AF moment which is $\sqrt{2}$ times smaller than the AF moment amplitude equal to 2 for the magnetisation given by formula (\ref{skyrmion}).
The experimentally determined magnetic moment amplitude shown in Table~\ref{magmom3} is $m_c=2.57\mu_{\rm B}$, which corresponds to the average moment $1.81\mu_{\rm B}$, the same value as for the constant moment helical structure (Table \ref{magmom1}). We present the evolution of the net topological charge to demonstrate qualitative effect of the field on topological charge density $w(x,y)$ and that for this type of structure the topological properties might persist up to FM saturation. This is indeed in accordance with the fact that the THE was observed up to the magnetic fields that saturate the magnetic moment to $0-1.9$ $\mu_{\rm B}$  per Mn-atom \cite{kanazawa2021} measured with the sample from the same batch. We note that the topological charge $Q$ can have values $Q=\pm1$, depending on the orientation of $m_f$. Since the THE is observed in the powder sample any sensitivity to the sign of $Q$ is not evident.

%
%
%

For the hedgehog structure given by formula (\ref{hedgehog}) the topological properties should also vanish in the ferromagnetic state which is achieved when the ferromagnetic component is larger than the amplitude $\sqrt{6}$. The experimental value of the amplitude for the 3+3 structure is the same as for the  3+2 one at 2.56~$\mu_{\rm B}$, implying the same FM saturation moment as for the meron-like (``skyrmion'') structure. The topological object can be visualised as shown in Fig.~\ref{yozh}. It shows a magnetisation distribution in a cube with edges $\pi/2$ around the centre $\pi/4,\pi/4,\pi/4$, where all components of the magnetisation become zero. The total solid angle spanned on the six cube faces gives $Q=-1$ in $4\pi$ units, similar to the skyrmion charge. There are eight objects like this in the magnetic unit cell for the parametrisation given by formula (\ref{hedgehog}) with the following positions (in units $\pi/4$) and charges Q: (-3,-3,-3),+1; (-3,-1,3),-1; (-1,-1,3),+1; (-1,3,-3),-1; (1,1,1),-1; (1,3,-1),+1; (3,-3,-1),-1; (3,-1,1),+1. The locations of these objects in the unit cell are illustrated in Fig.~SM4.
The trajectories of these objects in the magnetic field (called monopoles and anti-monopoles of emergent electromagnetic field) and their relation to the THE were calculated using realistic Hamiltonian with contributions from both the spin-orbit and spin-charge couplings \cite{Okumura}. Electric transport for three-dimensional skyrmion/monopole crystals was theoretically studied in~\cite{sk3_theory}.
%
%

\section{Conclusions}
\label{discon}
We have synthesised cubic monogermanide \mnge\ by several techniques and studied its crystal and magnetic structures by powder neutron diffraction. 
The propagation vectors of the magnetic structure are aligned with the [100] cubic axes and correspond to a
 length of magnetic modulation of about 30~\AA. 
We have found several maximal crystallographic symmetry magnetic structures that fit our diffraction data equally well.
 Among them there are two topological structures realised in the six-dimensional cubic magnetic superspace group (MSSG)  
198.3.206.1.m10.2 P2\_13.1'(a,0,0)00s(0,a,0)00s(0,0,a)00s and a five-dimensional orthorhombic one 19.2.29.2.m26.3 P2\_12\_12\_1.1'(0,b1,0)000s(0,0,g2)000s. 
The cubic  structure is of the hedgehog type whereby the magnetisation spatially modulates along all three dimensions, while  the orthorhombic one hosts meron-like objects located in a two-dimensional plane with total topological charge $|Q|=1$ per magnetic cell in infinitesimally small field  with the total modulation amplitudes 2.6~$\mu_B$. 
From the high resolution diffraction data we have found that the crystal structure is orthorhombic with small orthorhombic strain less than 0.16\%. The hedgehog magnetic structure is thus slightly distorted and in general might have more degrees of freedom in the subgroup MSSG 19.3.95.4.m26.4 P2\_12\_12\_1.1'(a1,0,0)000s(0,b2,0)000s(0,0,g3)000s.
Both 3$k$-cubic hedgehog and 2$k$-othorhombic meron structures can account for the topological Hall effect observed earlier in \mnge\ and should be preferable over the simple single-$k$ helical structure. The latter has been identified as MSSG 19.1.9.1.m26.2 P2\_12\_12\_1.1'(0b0)0s0s and can have constant moment helical configuration. 

We also report on a new combined mechanochemical and solid-state route to synthesise \mnge\ at ambient pressures and moderate temperatures, in addition to the traditional high pressure synthesis. The samples synthesised using the new approach
 have relatively small crystalline sizes of about 150~\AA. Nonetheless, the
 magnetic structures are the same with similar parameters as for the sample made by high pressure synthesis.  

\section*{A{\lowercase{cknowledgements}}}

This study was performed at Swiss neutron spallation SINQ. We thank the Swiss National Science foundation grants No. 200020-182536/1, 200021\_188706 and R'equip Grant No. 461 206021\_139082 and SNI Swiss Nanoscience Institute for financial support.
V.P. thanks V.~Markushin for the helpful discussions. 


%

\newpage

\begin{table*}

\caption{Magnetic structure parameters for MnGe for the different 3+1 models explained in section \ref{magstr}. $m_{jc},\,m_{js}$, $j=x,y,z$ are conventional crystallographic cosine and sine amplitudes of moment modulation
 along respective $x,y,z$-axes. These amplitudes correspond in pairs respectively to $m_j \cos(\alpha_j)$, $-m_j \sin(\alpha_j)$, $j=1,2,3$ for $m_j$ and $\alpha_j$ in formulae (\ref{mod1},\ref{mod3}). $M$ is the total amplitude of the modulation. (F) and (1) denote samples \mngef\ and \mngem, respectively. The goodness of fit~\cite{Fullprof} is also given for each model, $R$-factors are in percents. 
}\label{magmom1}

\begin{center}
\begin{tabular*}{\textwidth}{l @{\extracolsep{\fill}} llll}

model &  $m_{xc},\,m_{xs}$, $\mu_B$  &  $m_{yc},\,m_{ys}$, $\mu_B$  &  $m_{zc},\,m_{zs}$, $\mu_B$ &  $M, \mu_B$ \\ \hline 
3+1 (F)  SA AM          &  -2.5641, -0.1082  &  0.2967, 0.0480     & -0.4383, -0.0729    &      \\
\hline
3+1 (F)  SA helix           & -2.1362, -0.3394  & 0.1484, 0.0380 & -0.3210, 1.4200     &      \\
\hline
3+1 (F) AM-x     &  2.566(3), 0.33(6)  &            0, 0 & 0, 0           &   2.587(8)  \\
\multicolumn{2}{l}{$R_{wp}, R_{exp}, \chi^2, R_B$}& \multicolumn{3}{l}{3.64, 1.63, 5.01, 0.580 } \\ \hline
3+1 (F) AM-xz      &  2.566(3), 0 &          0,  0 & 0.48(6), 0   &   2.61(1)  \\
\multicolumn{2}{l}{$R_{wp}, R_{exp}, \chi^2, R_B$}& \multicolumn{3}{l}{3.63, 1.63, 4.97, 0.574 } \\ \hline
3+1 (F) helix     &   1.814(2), 0  &         0.   0 & 0, -1.814(2)      &    1.814(2)  \\
\multicolumn{2}{l}{$R_{wp}, R_{exp}, \chi^2, R_B$}& \multicolumn{3}{l}{  3.67, 1.63, 5.08, 0.625   } \\ \hline
3+1 (1) AM-xz      &   2.328(3), 0   &         0,   0 & 0.33(26), 0   &    2.35(4)  \\
\multicolumn{2}{l}{$R_{wp}, R_{exp}, \chi^2, R_B$}& \multicolumn{3}{l}{  7.59, 3.37, 5.07, 2.22   } \\ \hline

\end{tabular*}
\end{center}
\end{table*}

\begin{table*}

\caption{Magnetic structure parameters for MnGe for the different 3+3 and 3+2 models explained in section \ref{magstr}. See caption of Table~\ref{magmom1} for details.  The total moment amplitude, which is a sum over all $k$-vector components,  is $\sqrt{6}$ and 2 times larger than the component given for single $k$-vector for hedgehog and skyrmion structures, respectively. For the 3+2 structure $m_5$ and  $m_6$, are not given, because they are constrained to be equal to $m_1$ with phases $\alpha_5=\pi/2$, $\alpha_6=0$ in formula (\ref{mod2}).
}\label{magmom3}

\begin{center}
\begin{tabular*}{\textwidth}{l @{\extracolsep{\fill}} llll}

model   &  $m_{xc},\,m_{xs}$, $\mu_B$  &  $m_{yc},\,m_{ys}$, $\mu_B$  &  $m_{zc},\,m_{zs}$, $\mu_B$ &  $M, \mu_B$ \\ \hline 
3+3 (F)  SA     &    -0.8616, -0.0217  &  0.0028, 0.0711  & 0.1653, 1.2014    &      \\
\hline
3+3 (F) hedgehog &  1.048(1), 0   &0, 0   & 0, -1.048(1)                &  2.567(3)  \\
\multicolumn{2}{l}{$R_{wp}, R_{exp}, \chi^2, R_B$}& \multicolumn{3}{l}{3.67, 1.63, 5.08, 0.634   } \\ \hline
3+3 (1) hedgehog &  0.950(1), 0       &0, 0   & 0, -0.950(1)      &  2.327(3)  \\
\multicolumn{2}{l}{$R_{wp}, R_{exp}, \chi^2, R_B$}& \multicolumn{3}{l}{7.60, 3.37, 5.07, 2.22   } \\ \hline
3+3 (1) x &   1.344(2), 0.14(12)    &       0,  0 & 0, 0        &  2.328(3)    \\
\multicolumn{2}{l}{$R_{wp}, R_{exp}, \chi^2, R_B$}& \multicolumn{3}{l}{  7.60, 3.37, 5.07, 2.17   } \\ \hline
3+3 (F) xz &   1.42(5), 0  &      0,      0 & 0.28(3), 0.41(2)  &  2.56(6) \\
\multicolumn{2}{l}{$R_{wp}, R_{exp}, \chi^2, R_B$}& \multicolumn{3}{l}{3.62, 1.63, 4.95,  0.589   } \\ \hline
3+3 (F) x &  1.481(2), 0.19(3)   &0, 0   &0, 0   & 2.58(3) \\
\multicolumn{2}{l}{$R_{wp}, R_{exp}, \chi^2, R_B$}& \multicolumn{3}{l}{3.64, 1.63, 5.01, 0.569   } \\ \hline
3+2 (F) skyrmion &  1.283(1), 0    &0, 0   & 0, -1.283(1)      & 2.566(2)   \\
\multicolumn{2}{l}{$R_{wp}, R_{exp}, \chi^2, R_B$}& \multicolumn{3}{l}{3.67, 1.63, 5.08  , 0.643   } \\ \hline

\end{tabular*}
\end{center}
\end{table*}

\def\extgra{pdf}
\def\figsiz{\textwidth}

\begin{figure}
  \begin{center}
    \includegraphics[width=0.5\figsiz]{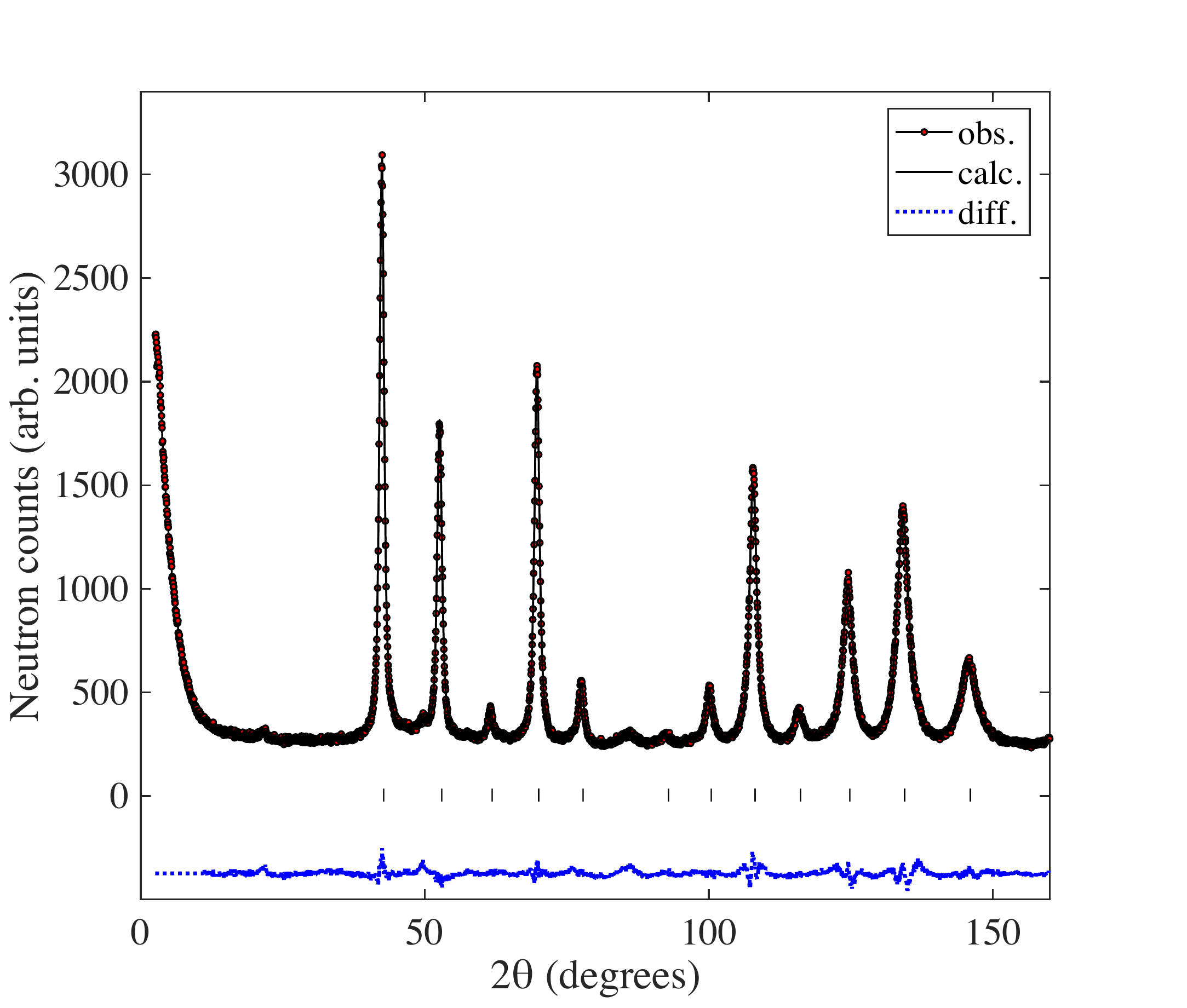} 
    \includegraphics[width=0.5\figsiz]{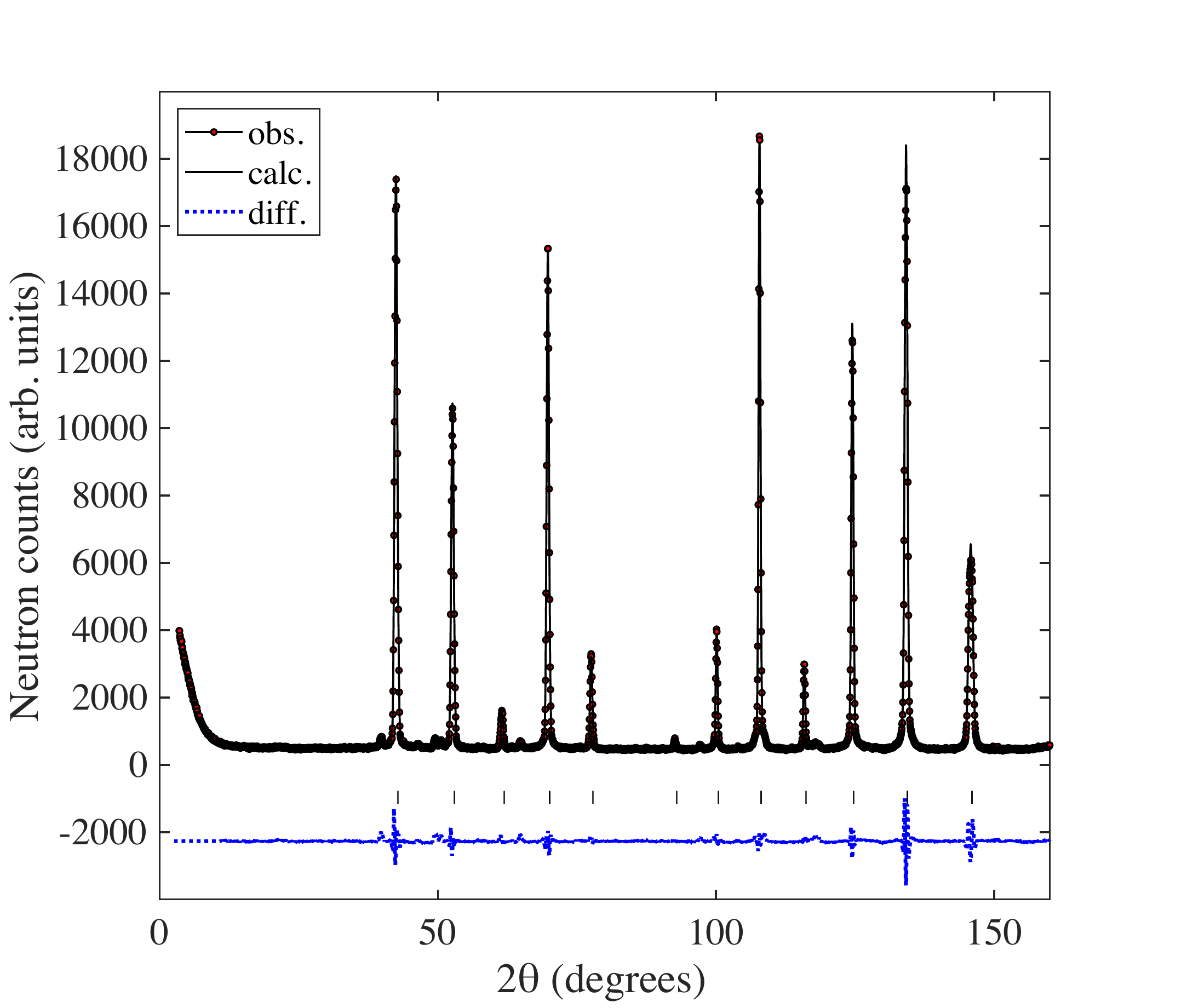} 
  \end{center}

  \caption{The Rietveld refinement pattern and difference plot of the
    neutron diffraction data for the samples  \mngem\ (top) and \mngef\ (bottom) at $T=300$~K
    measured at HRPT with the wavelength $\lambda=2.45$~\AA. The rows
    of tics show the Bragg peak positions. The difference between observed and calculated intensities is shown by the dotted blue line. The peak intensities in \mngef\ sample are larger due to narrower Bragg peaks.}
  \label{rt_diff}
\end{figure}

\begin{figure}
  \begin{center}
    \includegraphics[width=0.5\figsiz]{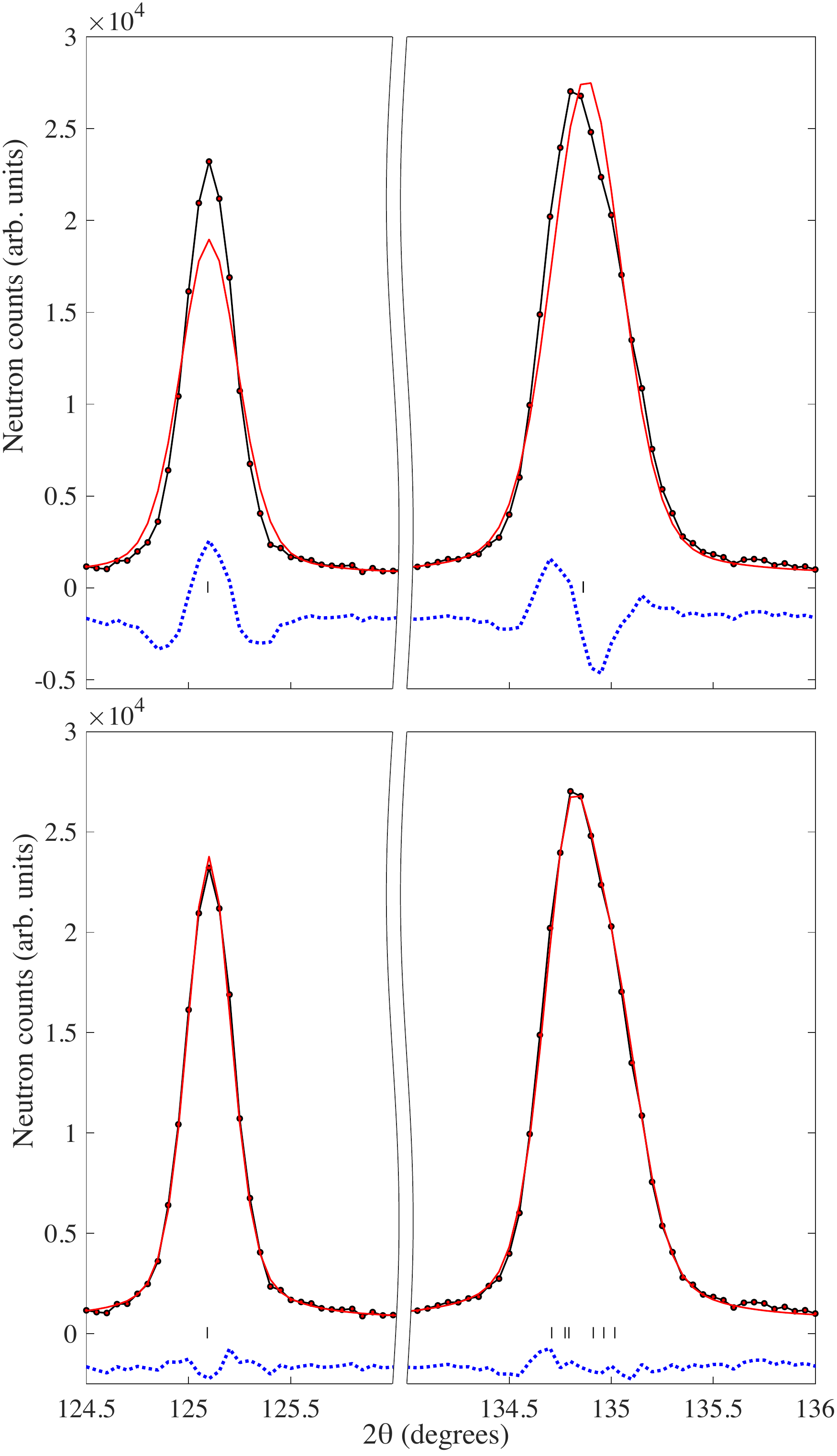} 
  \end{center}

  \caption{Fragment of the Rietveld refinement pattern and difference plot of the
    neutron diffraction data measured in HR-mode  at HRPT with the wavelength $\lambda=2.45$~\AA\ at T=1.8K. The upper plots are the fits in the parent cubic space group $P2_13$ with microstrain broadening. The bottom plots are for fits refinements done in the orthorhombic space group $P2_12_12_1$. The left Bragg peak is $(222)$ and is not split in both groups. One can see the shoulder of the right hand side peak (320) that is not possible to rationalise in the cubic group. The rows
    of ticks show the Bragg peak positions. The difference between observed and calculated intensities is shown by the dotted blue line.}
  \label{HR}
\end{figure}

\begin{figure}
  \begin{center}
    \includegraphics[width=0.5\figsiz]{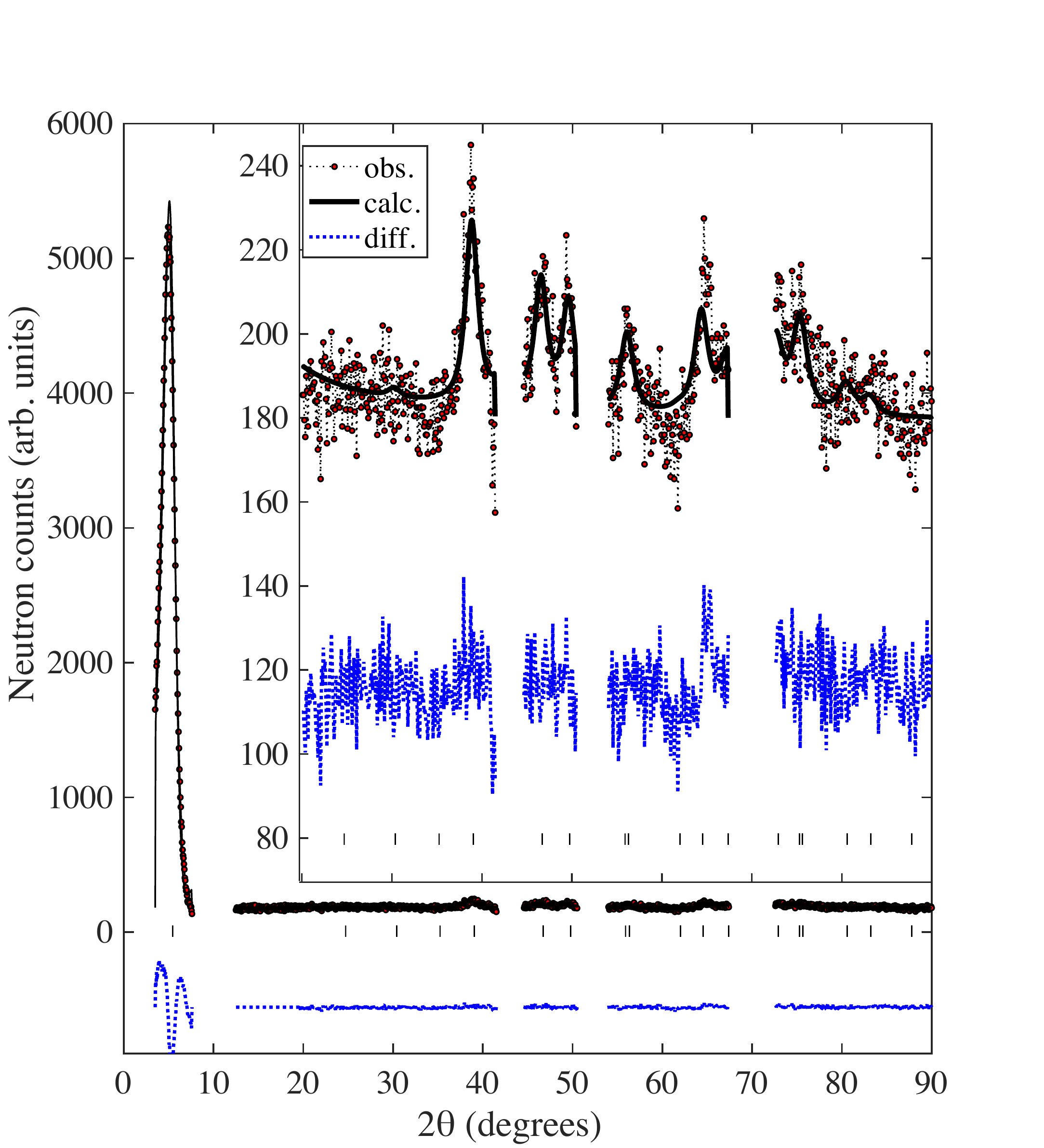} 
  \end{center}

  \caption{ Neutron powder diffraction pattern showing the difference between data measured with sample \mngem\ at $T=1.8$\,K and $T=300$\,K, with wavelength $\lambda=2.45$\,\AA. The inset shows the zoom in the large  $2\theta$ region with the same x-axis. The solid line shows the result of the fit to the magnetic model in 3D+1 MSSG. The row of vertical tick marks the the positions (hkl's) of the magnetic Bragg peaks. The difference between observed and calculated intensities is shown by the dotted blue line. See the text for details.}
  \label{mag_diff_01}
\end{figure}

\begin{figure}
  \begin{center}
    \includegraphics[width=0.5\figsiz]{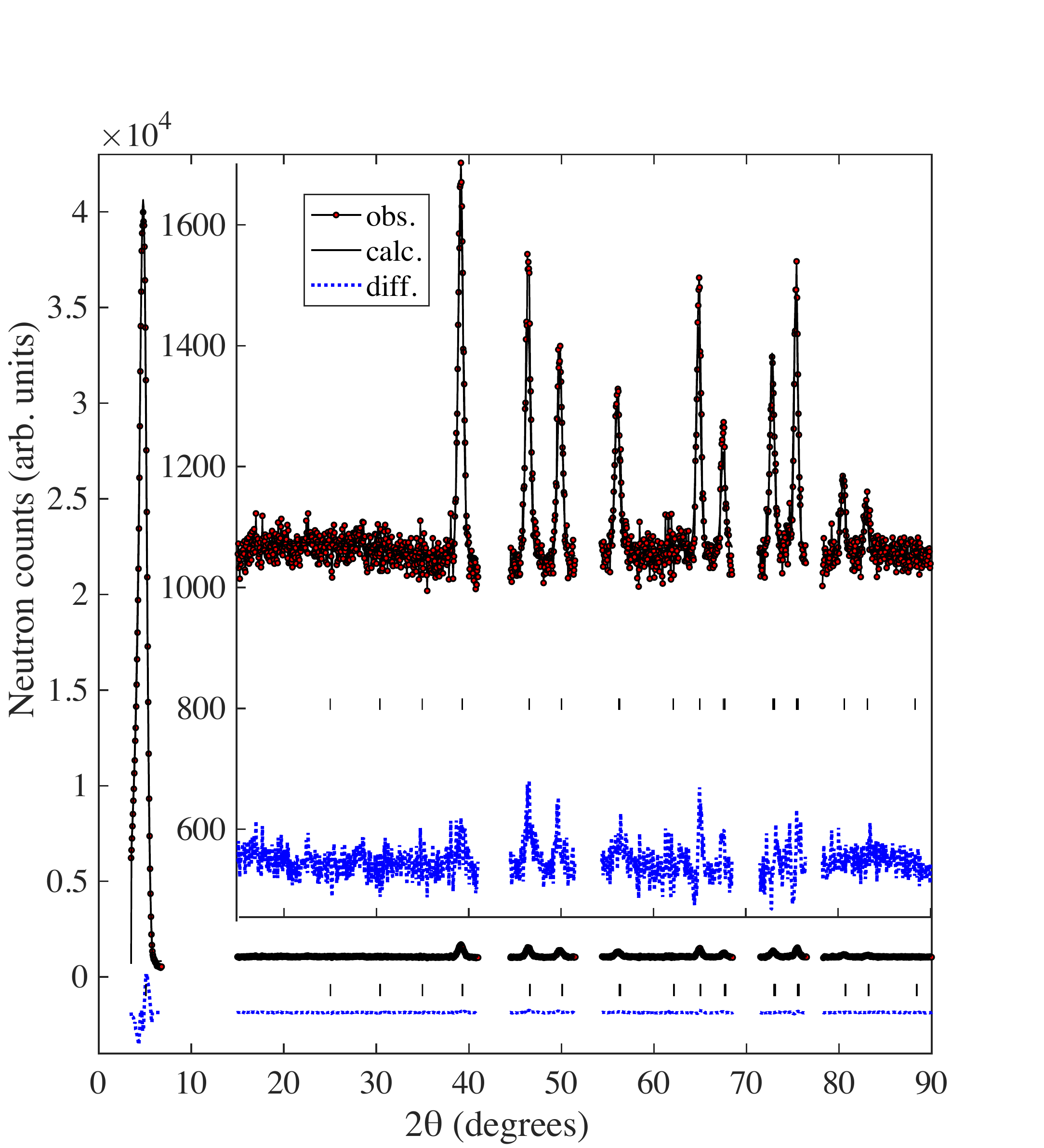} 
  \end{center}

  \caption{ Neutron powder diffraction pattern showing the difference between data measured with sample \mngef\ at $T=1.8$\,K and $T=300$\,K, with wavelength $\lambda=2.45$\,\AA. The inset shows the zoom in the large  $2\theta$ region with the same x-axis. The solid line shows the result of the fit to the magnetic model in 3D+1 MSSG. The row of vertical tick marks the the positions (hkl's) of the magnetic Bragg peaks. The difference between observed and calculated intensities is shown by the dotted blue line. See the text for details.}
  \label{mag_diff_F}
\end{figure}

\begin{figure}
  \begin{center}
    \includegraphics[width=0.5\figsiz]{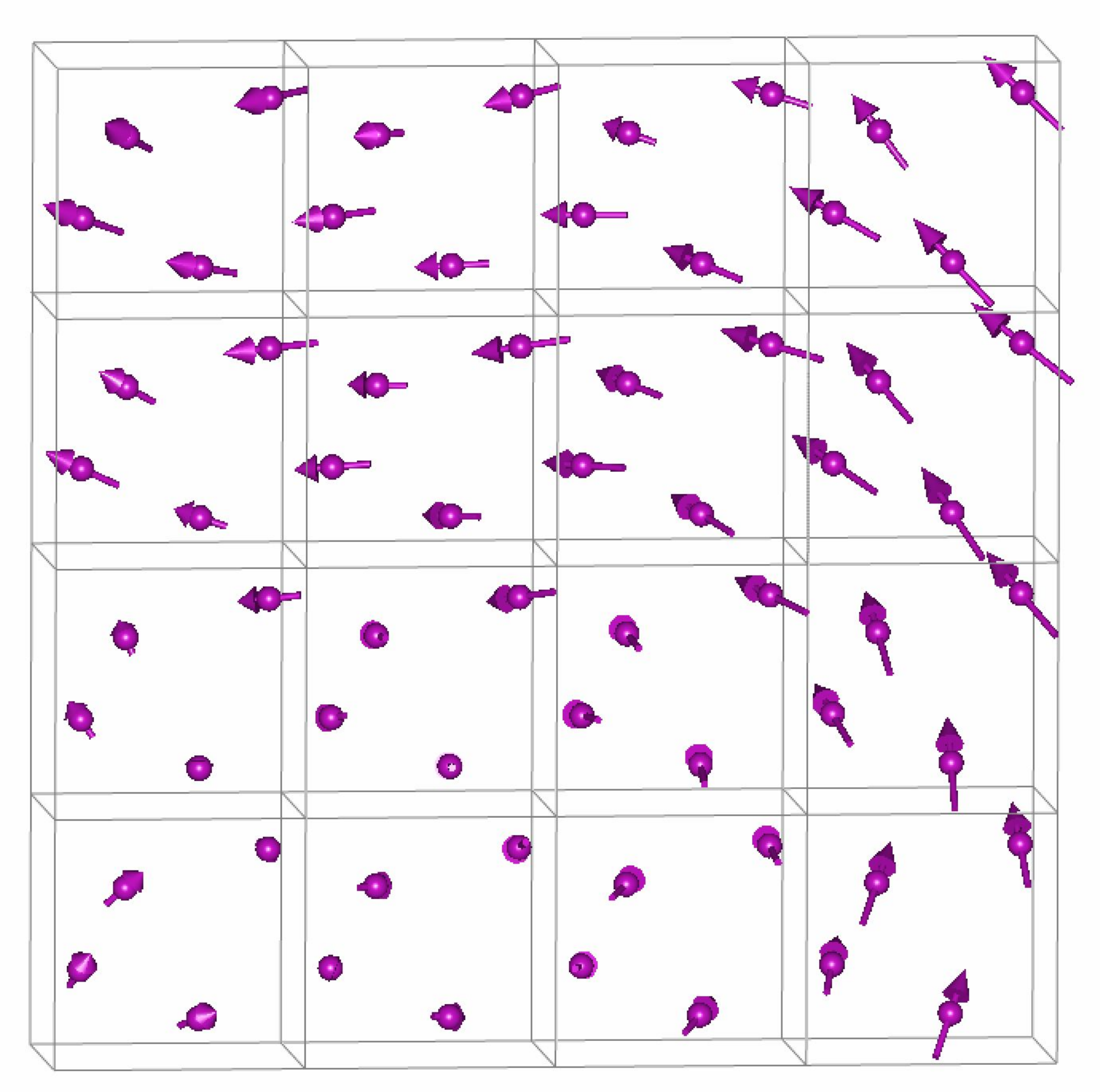} 
    \includegraphics[width=0.5\figsiz]{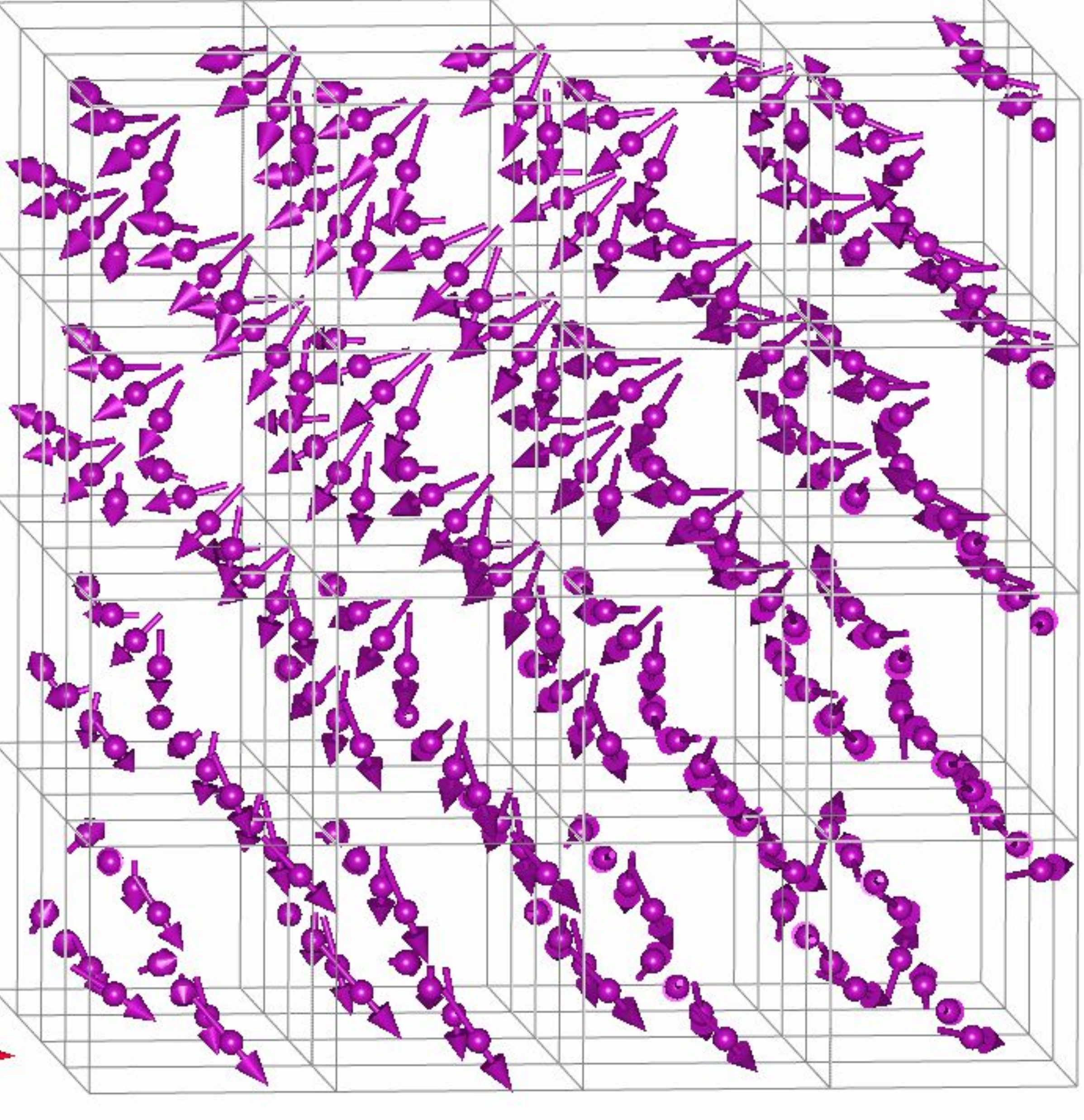} 
  \end{center}

  \caption{Magnetic structure in 3+3 hedgehog cubic model. 4x4x1 and 4x4x4 unit cells are shown in projection approximately along (001) axis. See table \ref{magmom3} for the parameters of the structure.}
  \label{mag_str_F3d4x4x1}
\end{figure}

\begin{figure}
  \begin{center}
    \includegraphics[width=0.5\figsiz]{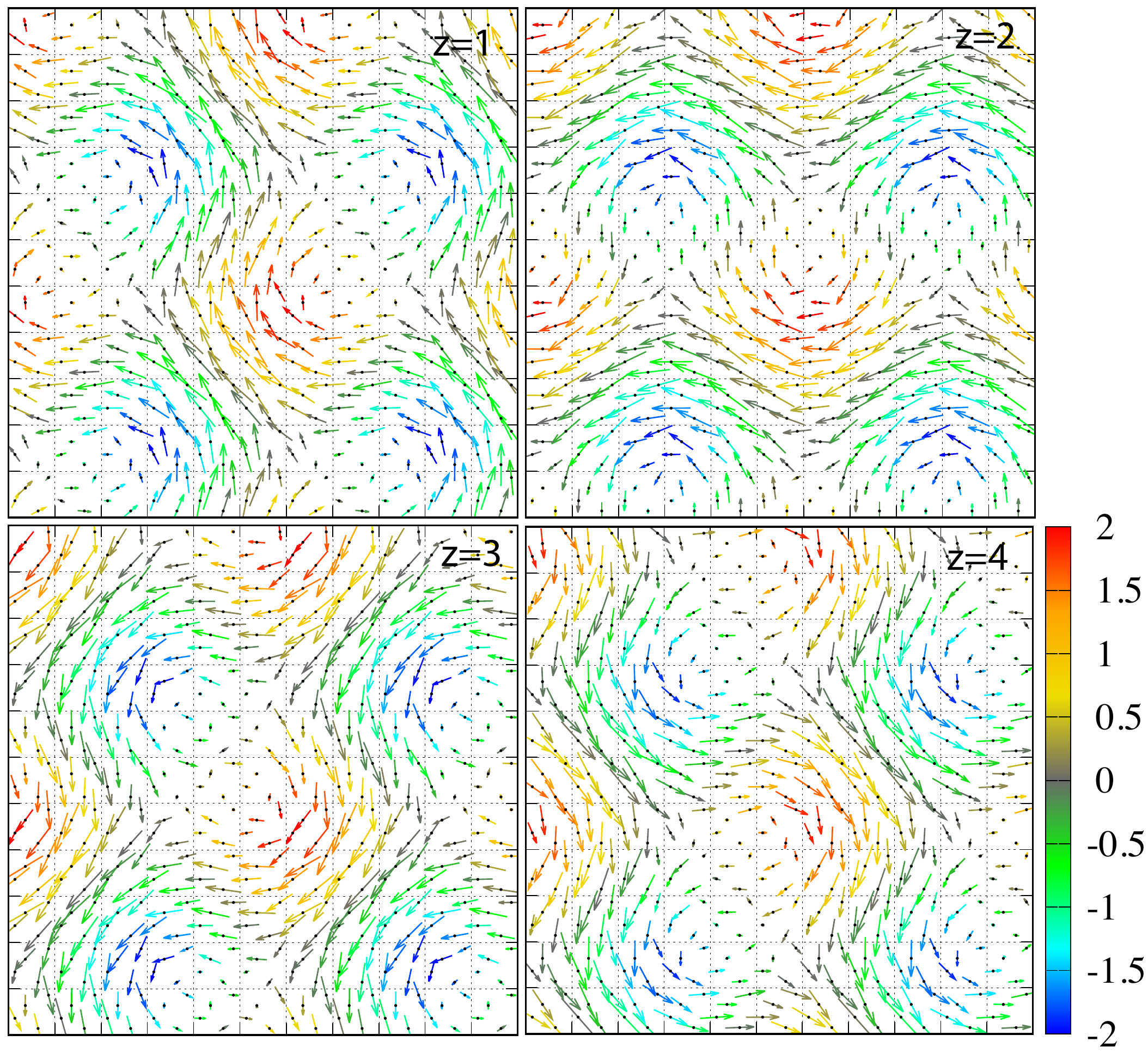} 
  \end{center}

  \caption{Magnetic structure in 3+3 cubic hedgehog model. 11x11x1 unit cells, corresponding to approximately 2x2 magnetic cells, are shown in projection along (001) axis for the z-layers indicated in the figure. The  $M_x,M_y$ components in the xy plane are the vector lengths, $M_z$ component is shown by colour. See table \ref{magmom3} for the parameters of the structure.}
  \label{mag_str_F3d11x11x1}
\end{figure}

\begin{figure}
  \begin{center}
    \includegraphics[width=0.5\figsiz]{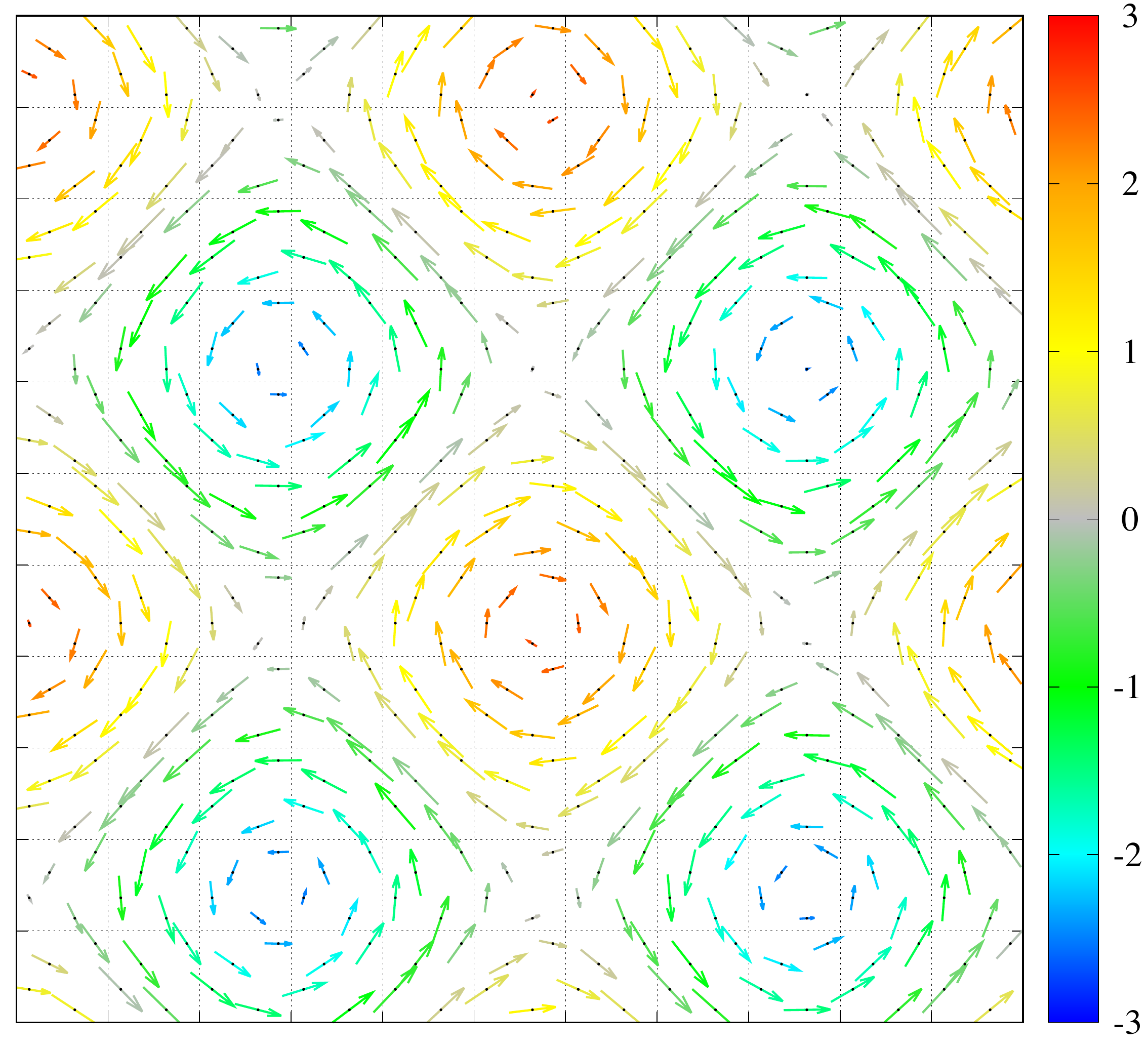} 

  \end{center}

  \caption{Magnetic structure in the 3+2 orthorhombic ``skyrmion'' model. First 11x11x1 unit cells, corresponding to approximately 2x2 magnetic cells, are shown in projection along the (001) axis. The  $M_x,M_y$ components in the xy plane are the vector lengths, $M_z$ component is shown by colour. See table \ref{magmom3} for the parameters of the structure.}
  \label{mag_str_F2d11x11x1}
\end{figure}

\begin{figure}
  \begin{center}
    \includegraphics[width=1\figsiz]{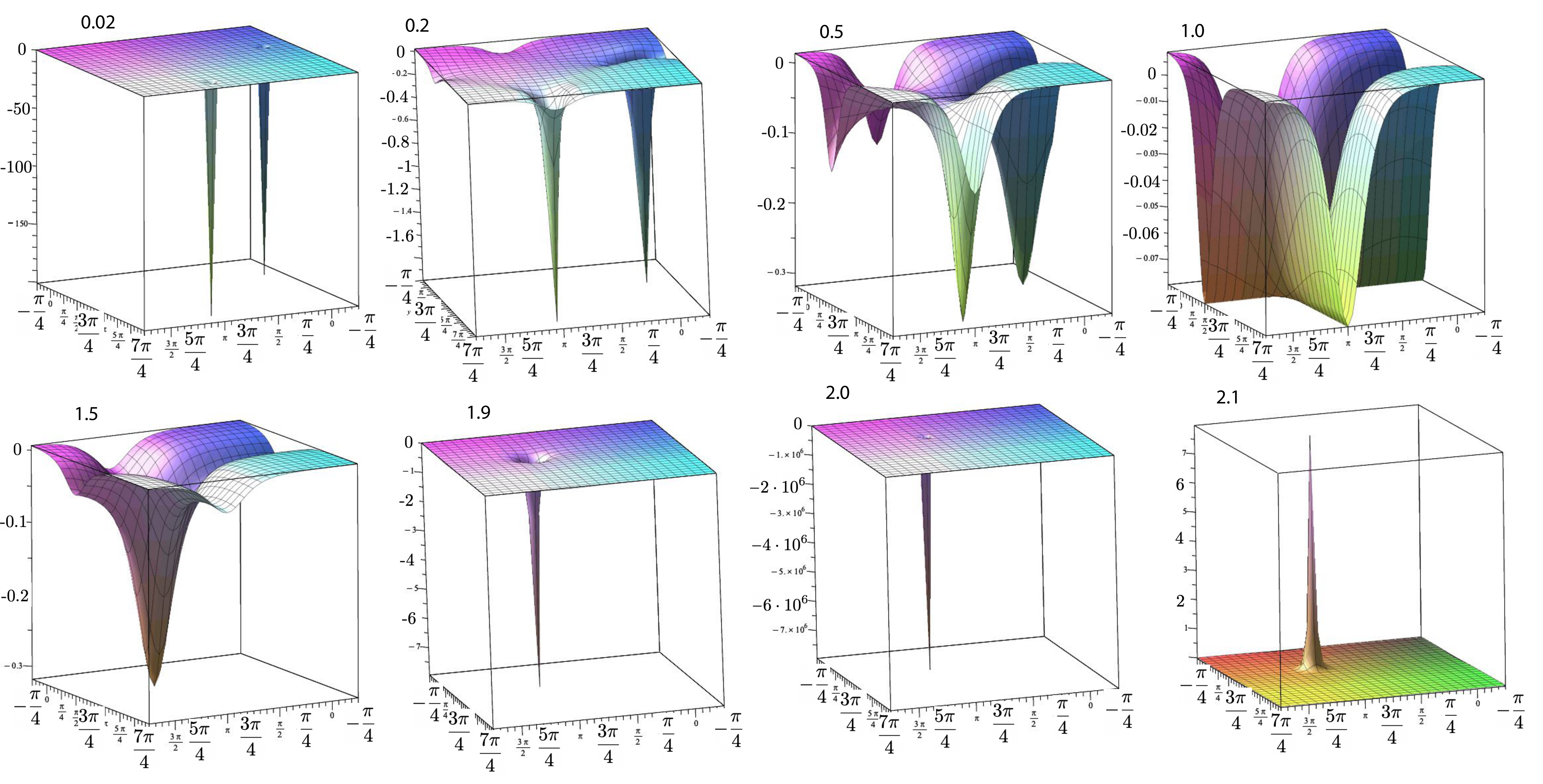} 

  \end{center}

  \caption{Density of topological charge $w(x,y)$ calculated using formula (\ref{top_idx}) for the magnetic structure in the orthorhombic 3+2 model given by formula (\ref{skyrmion}) (to avoid singularities the coefficient for cosine of the z-component was chosen to be 1.0001 
  with a ferromagnetic component along the $z$-axis $m_f=0.02, 0.2, 0.5, 1, 1.5, 1.9, 2.0, 2.1$. 
   One modulation period between $-\pi/4\, ..\, 2\pi-\pi/4$ is shown, corresponding to about 6 unit cells in Fig.~\ref{mag_str_F2d11x11x1}. Each peak carries topological  charge $Q=-1/2$ for infinitely small $m_f$.
   The total topological charge per cell is $Q=-1$ for $m_f\leq2$ and $Q=0$ for $m_f>2$.
  }
  \label{topo_den_mf}
\end{figure}

\begin{figure}
  \begin{center}
    \includegraphics[width=0.5\figsiz]{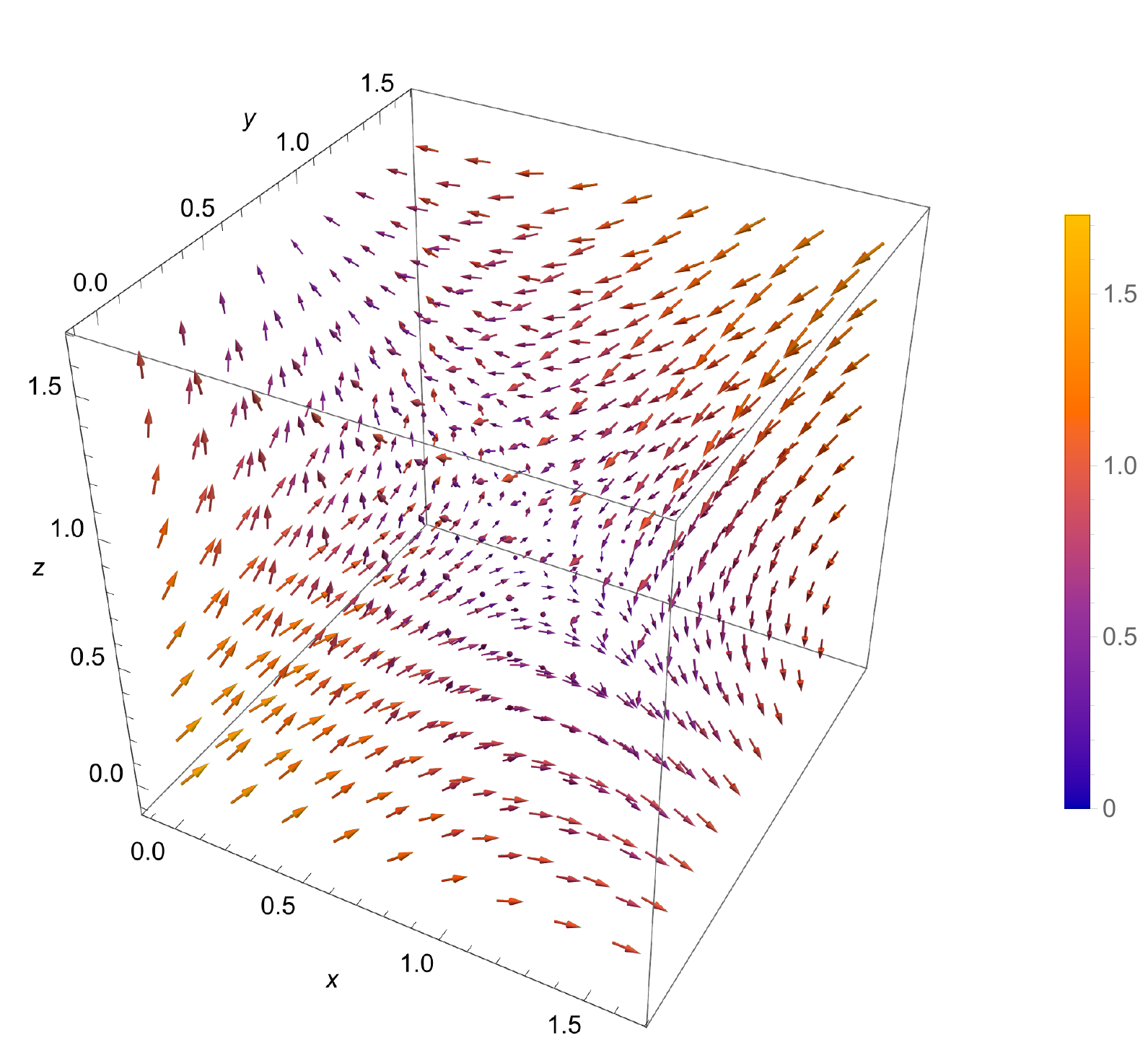} 
  \end{center}

  \caption{Fragment of magnetisation distribution for the cubic 3+3 hedgehog model
 given by (\ref{hedgehog}) in a cube with the edge $\pi/2$ around the centre $\pi/4,\pi/4,\pi/4$, where all magnetisation components become zero. The total solid angle spanned on the cube faces is $Q=-1$ in $4\pi$ units. The colour indicates the size of the magnetisation.}
  \label{yozh}
\end{figure}

\end{document}